\documentstyle[preprint,aps]{revtex}

\newcommand{\be}{\begin{equation}}
\newcommand{\ee}{\end{equation}}
\newcommand{\Dlt}{\Delta}
\newcommand{\dlt}{\delta}
\newcommand{\prt}{\partial}
\newcommand{\br}{{\bf r}}
\newcommand{\bk}{{\bf k}}
\newcommand{\bt}{\beta}
\newcommand{\vp}{\varphi}
\newcommand{\ep}{\varepsilon}
\newcommand{\al}{\alpha}
\newcommand{\ra}{\rightarrow}
\newcommand{\sgm}{\sigma}
\newcommand{\gm}{\gamma}
\newcommand{\om}{\omega}
\newcommand{\Gm}{\Gamma}
\newcommand{\dgr}{\dagger}
\newcommand{\lbd}{\lambda}

\tightenlines

\begin{document}

\draft

\title{Fluctuations of composite observables and stability of 
statistical systems} 

\author{V.I. Yukalov} 

\address{Institut f\"ur Theoretische Physik, \\ 
Freie Universit\"at Berlin, Arnimallee 14, D-14195 Berlin, Germany \\ 
and \\ 
Bogolubov Laboratory of Theoretical Physics, \\ 
Joint Institute for Nuclear Research, Dubna 141980, Russia}

\maketitle

\begin{abstract}

Thermodynamic stability of statistical systems requires that susceptibilities
be semipositive and finite. Susceptibilities are known to be related to the
fluctuations of extensive observable quantities. This relation becomes 
nontrivial, when the operator of an observable quantity is represented as 
a sum of operators corresponding to the extensive system parts. The 
association of the dispersions of the partial operator terms with the total 
dispersion is analyzed. A special attention is paid to the dependence of 
dispersions on the total number of particles $N$ in the thermodynamic limit. 
An operator dispersion is called thermodynamically normal, if it is 
proportional to $N$ at large values of the latter. While, if the dispersion 
is proportional to a higher power of $N$, it is termed thermodynamically 
anomalous. The following theorem is proved: The global dispersion of a 
composite operator, which is a sum of linearly independent self-adjoint 
terms, is thermodynamically anomalous if and only if at least one of the 
partial dispersions is anomalous, the power of $N$ in the global dispersion 
being defined by the largest partial dispersion. Conversely, the global 
dispersion is thermodynamically normal if and only if all partial dispersions 
are normal. The application of the theorem is illustrated by several examples 
of statistical systems. The notion of representative ensembles is formulated. 
The relation between the stability and equivalence of statistical ensembles 
is discussed.

\end{abstract}

\vskip 0.5cm

\pacs{05.40.-a, 05.70.-a, 05.70.Ce, 05.30.Jp}

\section{Introduction}

Stability of statistical systems and the fluctuations of observable 
quantities are known to be intimately related. The fluctuations can be 
characterized by the corresponding susceptibilities, such as specific 
heat, isothermal compressibility, or longitudinal magnetic susceptibility.
The susceptibilities are connected with the dispersions of the operators 
representing observable quantities. In what follows, we shall deal with 
the so-called {\it extensive} observables, whose averages are proportional 
to the total number of particles $N$, when $N$ is large [1,2]. The existence 
of the thermodynamic limit is assumed, when $N$ is asymptotically large, 
such that $N\ra\infty$.

Note that susceptibilities can also be connected with the fluctuations of 
intensive thermodynamic variables, such as pressure and temperature [3,4]. 
However, in this paper we shall consider only the fluctuations of extensive 
observables.

For stable statistical systems in equilibrium, the susceptibilities are 
positive and finite, which follows from their relations to the dispersions 
of the corresponding operator observables [5] or, on the general 
thermodynamic level, stems from the second law of thermodynamics [6]. The
susceptibilities may become divergent only at the points of second-order 
phase transitions, which, however, by definition, are the points of 
instability. Really, at the point of a phase transition, one phase becomes 
unstable, as a consequence, it transforms to another, stable, phase. After 
the phase transition has occurred, all susceptibilities in the stable phase 
go finite.

The fluctuations of extensive observables, related to the corresponding 
operator dispersions, can be classified onto two types, according to their 
dependence on the total number of particles $N$ in the given statistical 
system, when the number $N$ is large, such that $N\gg 1$. This implies 
that the thermodynamic limit is assumed. The fluctuations are called 
{\it thermodynamically normal}, when the related operator dispersion is 
proportional to $N$. Conversely, if the operator dispersion is proportional 
to $N^\al$, with $\al>1$, then the related fluctuations are termed {\it 
thermodynamically anomalous}.

The finiteness of susceptibilities in stable equilibrium systems means 
that the corresponding fluctuations are thermodynamically normal. 
Oppositely, the divergence of susceptibilities at the critical points 
shows that the fluctuations of the related extensive observables are 
thermodynamically anomalous. In a stable system, outside phase transition 
points, all susceptibilities are finite, which tells that the fluctuations 
of all extensive observables are thermodynamically normal.

It is worth warning the reader that thermodynamically normal or anomalous 
fluctuations have nothing to do with the normal, that is, Gaussian 
distributions. Thermodynamic normality or anomaly are the notions 
describing the thermodynamic behaviour of the related operator dispersions 
with respect to the total number of particles. In calculating the 
corresponding averages any quantum or classical probability measures, 
of arbitrary nature, can be employed.

In the present paper, general relations between the fluctuations of 
observables and the stability of statistical systems are studied. The 
emphasis is on the case, which is not a standard one, when the observable 
quantities are represented as sums of several terms, corresponding to 
macroscopic parts of the system. Then the relation between the fluctuations 
of the partial terms and the fluctuations of the global observables is not 
evident. A general theorem is rigorously proved, connecting the behaviour 
of fluctuations of global and partial observables. This theorem is briefly 
formulated in the Abstract and its mathematically rigorous formulation is 
given in Section III. The direct interrelation between the thermodynamic 
behaviour of fluctuations and stability is emphasized. It is also shown 
that the stability of statistical systems is intricately connected with 
the notions of symmetry breaking and ensemble equivalence.

\section{Fluctuations of observables and stability}

In quantum statistical mechanics, observable quantities are represented by
self-adjoint operators from the algebra of observables. As is explained in 
the Introduction, only extensive observables are considered in the paper.
Fluctuations of the observable quantities are characterized by the related 
operator dispersions. Let $\hat A$ be an operator representing an extensive 
observable quantity. Its dispersion is
\be
\label{1}
\Dlt^2(\hat A) \equiv \;  <\hat A^2>\; - \; <\hat A>^2 \; ,
\ee
where the angle brackets, as usual, denote statistical averaging.

The dispersions of the operators, representing extensive observables, are 
directly connected with the associated susceptibilities, which can be 
measured. Thus, the fluctuations of the Hamiltonian $H$, quantified by its
dispersion $\Dlt^2(H)$, define the specific heat
\be
\label{2}
C_V \equiv \frac{1}{N}\left ( \frac{\prt E}{\prt T}\right )_V =
\frac{\Dlt^2(\hat H)}{NT^2}\; ,
\ee
where $E\equiv<H>$ is internal energy, $N$ is the total number of particles
in the system of volume $V$, and $T$ is temperature. Here and in what follows, 
the Boltzman constant is set to unity, $k_B\equiv 1$. The fluctuations of the 
number of particles are described by the dispersion $\Dlt^2(\hat N)$ of the  
number-of-particle operator $\hat N$, yielding the isothermal compressibility
\be
\label{3}
\kappa_T \equiv -\;\frac{1}{V}\left ( \frac{\prt V}{\prt P}\right )_T =
\frac{\Dlt^2(\hat N)}{N\rho T}\; ,
\ee
in which $P$ is pressure, $N\equiv<\hat N>$, and $\rho\equiv N/V$ is the
average particle density. In magnetic systems, with the Zeeman interaction
$-\mu_0\sum_i{\bf B}\cdot{\bf S}_i$ of the operator spins ${\bf S}_i$ with
an external magnetic field ${\bf B}$, the fluctuations of the magnetization
$M_\al\equiv<\hat M_\al>$ are described by the dispersion $\Dlt^2(\hat M_\al)$
of the magnetization operator $\hat M_\al\equiv\mu_0\sum_{i=1}^N S_i^\al$,
which results in the longitudinal magnetic susceptibility
\be
\label{4}
\chi_\al \equiv \frac{1}{N}\left ( \frac{\prt M_\al}{\prt B_\al}\right )
= \frac{\Dlt^2(\hat M_\al)}{NT} \; .
\ee
In the notation, used above, $\mu_0=\hbar\gm_S$, with $\gm_S$ being the
gyromagnetic ratio for a particle of spin $S$. In what follows, we shall use
the system of units setting to unity the Planck constant $\hbar\equiv 1$.

The specific heat (2), isothermal compressibility (3), or magnetic
susceptibility (4) are the examples of the susceptibilities associated
with the fluctuations of observables. These thermodynamic characteristics
are readily measured in experiments. At the points of phase transitions,
the susceptibilities can diverge, since such points are the points of
instability. But for stable equilibrium system, the susceptibilities are
always positive and finite for all $N$, including the thermodynamic limit,
when $N\ra\infty$, $V\ra\infty$, so that $\rho\equiv N/V\ra const$. In
principle, it is admissible to imagine the situation, when a phase transition
occurs not merely at a point but in a finite region of a thermodynamic
variable [7], inside which region the system remains unstable and displays
a divergent susceptibility. Such a case, however, is quite marginal, and
rarely, if ever, happens for real statistical systems. In any event, as
soon as the phase transition is over, so that the system becomes stable,
all susceptibilities go finite.

The following picture summarizes the above consideration. The extensive 
observables of a statistical system are represented by Hermitian operators. 
The fluctuations of an observable, represented by an operator $\hat A$, 
are quantified by the operator dispersion $\Dlt^2(\hat A)$, whose ratio 
$\Dlt^2(\hat A)/N$ to the total number of particles characterizes the 
associated susceptibility. For a stable system, the latter must be 
semipositive and finite, while if it is divergent or negative, the system 
is unstable. This can be formulated as a {\it necessary stability condition}
\be
\label{5}
0 \leq \frac{\Dlt^2(\hat A)}{N} < \infty \; .
\ee
The ratio $\Dlt^2(\hat A)/N$ plays the role of a generalized susceptibility,
related to the operator $\hat A$. Examples of condition (5) are the stability
conditions on the specific heat (2), isothermal compressibility (3), and
magnetic susceptibility (4), according to which
\be
\label{6}
0 \leq C_V < \infty \; , \qquad 0 \leq \kappa_T < \infty\; , \qquad
0 \leq \chi_\al < \infty \; .
\ee
These thermodynamic characteristics are usually strictly positive at finite
temperature, becoming zero only at zero temperature.

In this way, the dispersion of the operator $\hat A$, representing an 
extensive observable, has to be proportional to the number of particles:
\be
\label{7}
\Dlt^2(\hat A) \propto N \; .
\ee
Then the dispersion is called {\it thermodynamically normal}. The thermodynamic
limit is assumed here, so that $N\gg 1$. When Eq. (7) is not satisfied, so that
$\Dlt^2(\hat A)\propto N^\al$ with $\al>1$, the dispersion is called {\it 
thermodynamically anomalous}. Respectively, the fluctuations of the related 
observable, characterized by the dispersion $\Dlt^2(\hat A)$, are termed  
thermodynamically normal, provided Eq. (7) is valid, and they are named 
thermodynamically anomalous if Eq. (7) does not hold.

In stable systems, the fluctuations of observables are always normal, and
the corresponding susceptibilities are finite. These susceptibilities can
be measured in experiment, either directly or through other measurable
quantities. For example, the isothermal compressibility can be measured
through the sound velocity
\be
\label{8}
s^2 \equiv \frac{1}{m}\left ( \frac{\prt P}{\prt\rho}\right )_T =
\frac{1}{m\rho\kappa_T} \; ,
\ee
where $m$ is the particle mass. The compressibility can also be found from
the central value of the structural factor
\be
\label{9}
S(0) = \frac{T}{ms^2} = \rho T \kappa_T \; .
\ee
And the structural factor
\be
\label{10}
S({\bf k}) = 1 +\rho \int [g(\br) -1] e^{-i{\bf k}\cdot\br}\; d\br \; ,
\ee
in which $g(\br)$ is the pair correlation function, can be measured in
scattering experiments.

\section{Theorem on total fluctuations}

In some cases, the operators of observables have the form of the sum
\be
\label{11}
\hat A = \sum_i \hat A_i
\ee
of self-adjoint terms $\hat A_i$. As has been stressed above, we consider 
here only extensive observables, such that the statistical average 
$<\hat A>$ is proportional to the total number of particles $N$, when 
the thermodynamic limit $N\ra\infty$ is implied. All parts $\hat A_i$ are 
assumed to have the same dimension as $\hat A$ and also to be the operators 
of extensive observables, so that $<\hat A_i>\propto N$. For example, 
$\hat A_1 =\hat K$ and $\hat A_2 =\hat W$ could be kinetic and potential 
energies for a system of $N$ particles. Then Eq. (11) would give the 
Hamiltonian $\hat H=\hat K+\hat W$. Or one can consider the operator of 
the number of particles $\hat N=\hat N_0+\hat N_1$ as a sum (11) composed 
of the operators of condensed particles, $\hat N_0$, and of noncondensed 
particles, $\hat N_1$, for a system with Bose-Einstein condensate. For 
each of the terms, one may consider partial fluctuations quantified by 
the dispersions $\Dlt^2(\hat A_i)$. Then of the principal interest is the 
problem how the partial dispersions $\Dlt^2(\hat A_i)$ are correlated with 
the total dispersion $\Dlt^2(\hat A)$? For instance, could it be that some 
of the partial dispersions are thermodynamically anomalous, while the total 
dispersion remains thermodynamically normal, so that the system as a total 
stays stable? The answer to such questions is given by the following 
theorem.

\vskip 2mm

{\bf Theorem}. Let the operator $\hat A$ of an extensive observable 
quantity be represented as a sum of linearly independent self-adjoint 
operators $\hat A_i$, being of the same dimension and also representing 
extensive observables, such that $<\hat A_i>\propto N$ in the thermodynamic 
limit. Then the global dispersion $\Dlt^2(\hat A)$ is thermodynamically 
anomalous, so that $\Dlt^2(\hat A)\propto N^\al$ with $\al>1$, if and 
only if at least one of the partial dispersions $\Dlt^2(\hat A_i)$ is 
thermodynamically anomalous. The power $\al$ in the dependence 
$\Dlt^2(\hat A)\propto N^\al$, as $N\ra\infty$, is defined by the 
largest power of all partial dispersions $\Dlt^2(\hat A_i)$. Conversely, 
the global dispersion $\Dlt^2(\hat A)$ is thermodynamically normal, such 
that $\Dlt^2(\hat A)\propto N$ in the thermodynamic limit, if and only 
if all partial dispersions $\Dlt^2(\hat A_i)$ are thermodynamically 
normal.

\vskip 2mm

{\bf Proof}. First, let us note that it is meaningful to consider only
linearly independent terms in the sum (11), since in the opposite case,
when some of the terms are linearly dependent, it is straightforward to
express one of them through the others, so that to reduce the number of
terms in sum (11). For  concreteness, in the following proof, the 
representatives of observables are called operators, which assumes the 
case of a quantum system. Of course, the same argumentation is valid for 
classical systems as well, for which one just has to replace the term 
"operator" by the term "classical random variable".

The dispersion for the operator sum (11) can be written as
\be
\label{12}
\Dlt^2(\hat A) = \sum_i \Dlt^2(\hat A_i) +
2 \sum_{i<j} {\rm cov}(\hat A_i,\hat A_j)\; ,
\ee
where the covariance
\be
\label{13}
 {\rm cov}(\hat A_i,\hat A_j) \equiv \frac{1}{2}<\hat A_i\hat A_j +
\hat A_j\hat A_i> - <\hat A_i><\hat A_j>
\ee
is employed. The latter enjoys the symmetry property
$$
 {\rm cov}(\hat A_i,\hat A_j) =  {\rm cov}(\hat A_j,\hat A_i) \; .
$$
The dispersions are, by definition, semipositive, while the covariances
can be positive as well as negative.

It is sufficient to prove the theorem for the sum of two operators, when
\be
\label{14}
\Dlt^2(\hat A_i + \hat A_j) = \Dlt^2(\hat A_i) + \Dlt^2(\hat A_j) +
2 {\rm cov}(\hat A_i,\hat A_j) \; .
\ee
This follows from the simple fact that any sum of terms more than two can
always be redefined as a sum of two new terms. We assume that in Eq. (14),
where $i\neq j$, both terms are operators but not classical functions. If
one of the terms were just a classical function, then we would have a
trivial equality
$$
\Dlt^2(\hat A_i +const) =\Dlt^2(\hat A_i) \; ,
$$
with the left-hand and right-hand sides being simultaneously either 
thermodynamically normal or anomalous.

The elements
\be
\label{15}
\sgm_{ij} \equiv{\rm cov}(\hat A_i, \hat A_j) \; ,
\ee
having the properties $\sgm_{ii}=\Dlt^2(\hat A_i)\geq 0$ and
$\sgm_{ij}=\sgm_{ji}$, form the covariance matrix $[\sgm_{ij}]$. This
matrix is symmetric. For a set of arbitrary real-valued numbers $x_i$, with
$i=1,2,\ldots,n$, where $n$ is an integer, one has
\be
\label{16}
 < \left [ \sum_{i=1}^n \left (\hat A_i -\; < \hat A_i>
\right ) x_i \right ]^2  > = \sum_{i,j=1}^n \sgm_{ij} x_i x_j
\geq 0 \; .
\ee
The right-hand side of equality (16) is a semipositive quadratic form.
The theory of quadratic forms [8] tells us that a quadratic form is
semipositive if and only if all principal minors of its coefficient matrix
are non-negative. Thus, the sequential principal minors of the covariance
matrix $[\sgm_{ij}]$, with $i,j=1,2,\ldots,n$, are all non-negative. In
particular,
$$
\sgm_{ii}\sgm_{jj} - \sgm_{ij}\sgm_{ji} \geq 0 \; .
$$
This, because of the symmetry $\sgm_{ij}=\sgm_{ji}$, takes the form
$$
\sgm_{ij}^2 \leq \sgm_{ii}\sgm_{jj} \; .
$$
Hence, the correlation coefficient
\be
\label{17}
\lbd_{ij} \equiv \frac{\sgm_{ij}}{\sqrt{ \sgm_{ii}\sgm_{jj} }}
\ee
possesses the property
$$
\lbd_{ij}^2 \leq 1 \; .
$$

The equality $\lbd_{ij}^2=1$ holds true if and only if $\hat A_i$ and
$\hat A_j$ are linearly dependent. The sufficient condition is evident, since
if $\hat A_j=a+b\hat A_i$, with $a$ and $b$ being any real numbers, then
$\sgm_{ij}=b\sgm_{ii}$ and $\sgm_{jj}=b^2\sgm_{ii}$, thence $\lbd_{ij}=b/|b|$,
from where $\lbd^2_{ij}=1$. To prove the necessary condition, let us assume
that $\lbd^2_{ij}=1$. Therefore $\lbd_{ij}=\pm 1$. Let us consider the
dispersion
$$
\Dlt^2\left ( \frac{\hat A_i}{\sqrt{\sgm_{ii}}} \pm
\frac{\hat A_j}{\sqrt{\sgm_{jj}}} \right ) = 2(1 \pm\lbd_{ij}) \geq 0 \; .
$$
The value $\lbd_{ij}=1$ is possible then and only then, when
$$
\Dlt^2\left ( \frac{\hat A_i}{\sqrt{\sgm_{ii}}} -
\frac{\hat A_j}{\sqrt{\sgm_{jj}}} \right ) = 0\; .
$$
The dispersion can be zero if and only if
$$
\frac{\hat A_i}{\sqrt{\sgm_{ii}}} -
\frac{\hat A_j}{\sqrt{\sgm_{jj}}} = const \; ,
$$
which implies that the operators $\hat A_i$ and $\hat A_j$ are linearly
dependent. In the same way, the value $\lbd_{ij}=-1$ is possible if and
only if
$$
\Dlt^2\left ( \frac{\hat A_i}{\sqrt{\sgm_{ii}}} +
\frac{\hat A_j}{\sqrt{\sgm_{jj}}} \right ) =0 \; .
$$
And this is admissible then and only then, when
$$
\frac{\hat A_i}{\sqrt{\sgm_{ii}}} +
\frac{\hat A_j}{\sqrt{\sgm_{jj}}} = const \; ,
$$
which again means the linear dependence of the operators $\hat A_i$
and $\hat A_j$. As far as these operators, by assumption, are linearly
independent, one has
\be
\label{18}
\lbd_{ij}^2 < 1\; .
\ee
This inequality is equivalent to
$$
\sgm_{ij}^2 < \sgm_{ii}\sgm_{jj} \; ,
$$
which, employing notation (15), becomes
\be
\label{19}
\left | {\rm cov}(\hat A_i,\hat A_j)\right |^2 < \Dlt^2(\hat A_i)
\Dlt^2(\hat A_j) \; .
\ee

Now, equality (14) can be represented as
\be
\label{20}
\Dlt^2(\hat A_i + \hat A_j) = \sgm_{ii} + \sgm_{jj} +
2\lbd_{ij}\sqrt{\sgm_{ii}\sgm_{jj}} \; ,
\ee
where, as is shown above, $|\lbd_{ij}|<1$. Altogether there can occur no
more than four following cases. First, both partial dispersions
$\sgm_{ii}=\Dlt^2(\hat A_i)$ and $\sgm_{jj}=\Dlt^2(\hat A_j)$ are normal,
so that $\sgm_{ii}\propto N$ and  $\sgm_{jj}\propto N$. Then, from Eq.
(20) it is obvious that the total dispersion $\Dlt^2(\hat A_i + \hat A_j)
\propto N$ is also normal. Second, one of the partial dispersions, say 
$\sgm_{ii}\propto N$, is normal, but another one is anomalous, $\sgm_{jj}
\propto N^\al$, with $\al>1$. From Eq. (20), using the inequality 
$(1+\al)/2<\al$, one has $\Dlt^2(\hat A_i + \hat A_j)\propto N^\al$. 
That is, the total dispersion is anomalous, with the same power $\al$ 
as $\sgm_{jj}$. Third, both partial dispersions are anomalous, such that 
$\sgm_{ii}\propto N^{\al_i}$ and $\sgm_{jj}\propto N^{\al_j}$ with different 
powers, say $1<\al_i<\al_j$. Then Eq. (20), with taking account of the 
inequality $(\al_i+\al_j)/2<\al_j$, shows that $\Dlt^2(\hat A_i + \hat A_j)
\propto N^{\al_j}$. Hence, the total dispersion is also anomalous, with the 
power $\al_j$ of the largest partial dispersion $\sgm_{jj}$. Fourth, both 
partial dispersions are anomalous, $\sgm_{ii}\propto c_i^2N^\al$ and 
$\sgm_{jj}\propto c_j^2N^\al$, where $c_i>0$ and $c_j>0$, with the same 
power $\al$. In that case, Eq. (20) yields $\Dlt^2(\hat A_i+\hat A_j)=
c_{ij}N^\al$, where
$$
c_{ij} \; \propto \; (c_i-c_j)^2 + 2 c_i c_j (1 +\lbd_{ij}) > 0 \; ,
$$
which is strictly positive in view of inequality (18). That is, the total
dispersion is anomalous, having the same power $\al$ of $N$ as both partial
dispersions. After listing all admissible cases, we see that the total
dispersion is anomalous if and only if at least one of its partial
dispersions is anomalous, with the power of $N$ of the total dispersion
being equal to the largest power of partial dispersions. Conversely, the
total dispersion is normal if and only if all its partial dispersions are
normal. This concludes the proof of the theorem.

\vskip 2mm

This theorem was, first, announced, without proof, in Ref. [9]. The proof,
presented above, is rather general, being valid for arbitrary operators
and statistical systems. The theorem can be applied to any system. For
instance, this can be a multicomponent system, where the index $i$ in Eq.
(11) enumerates the components. In recent years, much attention is given
to systems with Bose-Einstein condensate (see review articles [10--12]). 
The problem of fluctuations in such systems has received a great deal of
attention, with a number of papers claiming the existence of anomalous
fluctuations everywhere below the condensation point (see discussion in
Ref. [13]). In the following sections, the examples of Bose-condensed
systems will be considered. In addition to being naturally separated into
the condensed and noncondensed parts, Bose systems can also display the
coexistence of several coherent topological modes [14--23]. Another
possibility is the coexistence of atoms in several internal states,
which, e.g., has been studied in collective Raman scattering [24].

\section{Ideal Bose gas}

The uniform ideal Bose gas below the condensation temperature is known to
exhibit anomalous number-of-particle fluctuations [25,26]. Here, this case
will be briefly recalled for the purpose of illustrating the above
theorem.

The condensation temperature of the ideal uniform Bose gas is
\be
\label{21}
T_c = \frac{2\pi}{m}\left [ \frac{\rho}{\zeta(3/2)}\right ]^{2/3} \; ,
\ee
where $\zeta(3/2)\cong 2.612$. Below this temperature, the number-of-particle
operator is the sum
\be
\label{22}
\hat N =\hat N_0 + \hat N_1
\ee
of the terms corresponding to condensed and noncondensed particles,
respectively,
$$
\hat N_0 = a_0^\dgr a_0 \; , \qquad
\hat N_1 = \sum_{k\neq 0} a_k^\dgr a_k \; ,
$$
where $a_k^\dgr$ and $a_k$ are the creation and annihilation operators 
of Bose particles with momentum $k$.

The dispersion for the total number-of-particle operator $\hat N$ can be
calculated by means of the derivative over the chemical potential $\mu$,
so that
\be
\label{23}
\Dlt^2(\hat N) = T \; \frac{\prt N}{\prt\mu} \qquad (\mu\ra \; -0) \; .
\ee
The average number of particles $N=<\hat N>$ is given by the sum
\be
\label{24}
N = N_0 + N_1
\ee
of condensed,
\be
\label{25}
N_0 \equiv\; <a_0^\dgr a_0>\; =\; \left ( e^{-\bt\mu} - 1
\right )^{-1} \; ,
\ee
and noncondensed,
\be
\label{26}
N_1 \equiv \; < \hat N_1>\; = \; \frac{N}{\rho\lbd_T^3}\; g_{3/2}
\left ( e^{\bt\mu}\right ) \; ,
\ee
particles, where $\mu\ra-0$,
$$
\lbd_T \equiv \sqrt{\frac{2\pi}{mT}}\; , \qquad \bt \equiv 
\frac{1}{T}\; ,
$$
and the Bose-Einstein function is
$$
g_n(z) \equiv \frac{1}{\Gm(n)}\; \int_0^\infty \;
\frac{zu^{n-1}}{e^u-z}\; du \; .
$$
Let us stress that the terms $\hat N_0$ and $\hat N_1$ in the sum (23) 
are linearly independent. Differentiating the sum (24), one has the total 
dispersion
\be
\label{27}
\Dlt^2(\hat N) = \Dlt^2(\hat N_0) + \Dlt^2(\hat N_1) \; ,
\ee
with the partial dispersions
$$
\Dlt^2(\hat N_0) = T\; \frac{\prt N_0}{\prt\mu} \; , \qquad
\Dlt^2(\hat N_1) = T \; \frac{\prt N_1}{\prt\mu} \; .
$$
From Eqs. (25) and (26), we find the dispersion for condensed particles,
\be
\label{28}
\Dlt^2(\hat N_0) = N_0(1+ N_0) \; ,
\ee
and for noncondensed particles,
\be
\label{29}
\Dlt^2(\hat N_1) = \frac{N}{\rho\lbd_T^3}\;
g_{1/2}\left ( e^{\bt\mu}\right ) \; ,
\ee
where $\mu\ra-0$. As far as the existence of Bose-Einstein condensate 
presupposes that the number of condensed particles $N_0$ is macroscopic, 
that is, proportional to $N$, then from Eq. (28) and the relation $N_0
\propto N\gg 1$, we have $\Dlt^2(N_0)\propto N^2$. Expression (29) in the 
thermodynamic limit possesses an infrared divergence caused by the integral
$$
g_{1/2}(1) \propto \frac{1}{\sqrt{\pi}}\; \int_{u_{min}}^\infty \;
\frac{du}{u^{3/2}} \; ,
$$
in which
$$
u_{min} =\frac{k^2_{min}}{2mT}\; , \qquad k_{min}\propto \frac{1}{L} \; ,
$$
with $L\propto V^{1/3}$. Consequently, $g_{1/2}(1)\propto L/\lbd_T$.
Thus,
dispersion (29) diverges at finite temperatures as
\be
\label{30}
\Dlt^2(\hat N_1) \propto (mT)^2 V^{4/3} \; .
\ee
In this way, both dispersions for the number-of-particle operators of
condensed as well as noncondensed particles are anomalous:
$$
\Dlt^2(\hat N_0)\propto N^2 \; , \qquad \Dlt^2(\hat N_1) \propto
N^{4/3} \; .
$$
As a result, the total dispersion (27) is also anomalous, $\Dlt^2(\hat N)
\propto N^2$, with the power of $N$ given by $\Dlt^2(\hat N_0)$.

The anomalous dispersion $\Dlt^2(\hat N)$ leads, according to Eq. (3),
to the divergence of the isothermal compressibility, as $\kappa_T\propto N$,
everywhere below $T_c$, except $T=0$. But the system with a divergent
compressibility is not stable. Therefore, the ideal uniform Bose gas below
the condensation temperature (21) is a pathological object, being unstable
in the whole region $0<T\leq T_c$. In other words, such a gas does not
exist as a stable statistical system [13].

It is worth emphasizing that the anomalous fluctuations of the condensate
can be cured by breaking gauge symmetry as will be explained below. However 
the fluctuations of noncondensed particles remain anomalous, with the 
dispersion $\Dlt^2(\hat N_1)\propto N_1^{4/3}$ in both ensembles, grand 
canonical as well as canonical [25,26]. Therefore, the instability of the 
ideal uniform Bose gas below $T_c$ is not an artifact caused by the choice 
of an ensemble, but a property peculiar to this system.

\section{Interacting Bose gas}

There exists a popular myth that the number-of-particle fluctuations of
noncondensed particles in an interacting Bose gas below $T_c$ remain
anomalous, corresponding to the dispersion $\Dlt^2(\hat N_1)\propto N^{4/3}$,
of the same type as that for the ideal Bose gas (see discussion in Ref. [13]).
If this were true, then according to the theorem of Section 3, the total
dispersion $\Dlt^2(\hat N)$ would also be anomalous, with the power of $N$
not smaller than $4/3$. This would imply that the isothermal compressibility
diverges at least as $\kappa_T\propto N^{1/3}$. Hence the system as a whole
would be unstable. In turn, this would mean that there are no stable
statistical systems with Bose-Einstein condensate. Such a conclusion, of
course, would be too radical, because of which it is necessary to reconsider
the procedure of calculating the number-of-particle dispersions for
Bose-condensed systems.

Let us consider a weakly interacting Bose gas at low temperatures, when
the Bogolubov theory [27--29] is applicable. The main points of this theory
are as follows. One starts with the standard Hamiltonian
\be
\label{31}
H = \int \psi^\dgr(\br)\left ( -\; \frac{\nabla^2}{2m} - \mu \right )
\psi(\br)\; d\br + \frac{1}{2}\; \int \psi^\dgr(\br)\psi^\dgr(\br')
\Phi(\br-\br')\psi(\br')\psi(\br)\; d\br d\br'
\ee
in terms of the Bose field operators $\psi(\br)$ and $\psi^\dgr(\br)$. The
interaction potential is assumed to be symmetric, such that $\Phi(-\br)=
\Phi(\br)$, and soft, allowing for the Fourier transformation
$$
\Phi(\br) = \frac{1}{V}\; \sum_k \Phi_k e^{i{\bf k}\cdot\br} \; ,
\qquad \Phi_k = \int \Phi(\br) e^{-i{\bf k}\cdot{\bf r}} \; d\br \; .
$$

The condensate is separated by means of the Bogolubov shift
\be
\label{32}
\psi(\br) = \psi_0 +\psi_1(\br) \; ,
\ee
in which
\be
\label{33}
\psi_0 = \frac{a_0}{\sqrt{V}} \; , \qquad \psi_1(\br) = \sum_{k\neq 0}
a_k\vp_k(\br) \; ,
\ee
and, keeping in mind a uniform system, the expansion is over the plane
waves $\vp_k(\br)=e^{i{\bf k}\cdot\br}/\sqrt{V}$. The gauge symmetry of
Hamiltonian (31) is broken by setting $a_0=\sqrt{N_0}$. Assuming that $N_0
\approx N$, one omits from the total Hamiltonian the terms of the third and
fourth order with respect to the operators $a_k$ of noncondensed particles,
where $k\neq 0$. Retaining only the terms up to the second order in $a_k$,
one gets the quadratic Hamiltonian
\be
\label{34}
H_2 = \frac{1}{2}\; N\rho \Phi_0 + \sum_{k\neq 0} \om_k a_k^\dgr a_k -
\mu N + \frac{1}{2}\; \sum_{k\neq 0} \Dlt_k\left ( a_k^\dgr a_{-k}^\dgr
+ a_{-k} a_k\right ) \; ,
\ee
in which the notation for the quantities
\be
\label{35}
\om_k \equiv \frac{k^2}{2m} + \rho (\Phi_0 +\Phi_k) - \mu
\ee
and
\be
\label{36}
\Dlt_k \equiv \rho\Phi_k
\ee
is employed.

The quadratic Hamiltonian (34) is diagonalized by means of the Bogolubov
canonical transformation
$$
a_k = u_k b_k + v_{-k}^* b_{-k}^\dgr \; ,
$$
in which
$$
u_k^2 - v_k^2 =1 \; , \qquad u_k v_k = -\; \frac{\Dlt_k}{2\ep_k}\; ,
$$
$$
u_k^2 = \frac{\sqrt{\ep_k^2+\Dlt^2_k}+\ep_k}{2\ep_k}=
\frac{\om_k+\ep_k}{2\ep_k}\; , \qquad
v_k^2 = \frac{\sqrt{\ep_k^2+\Dlt^2_k}-\ep_k}{2\ep_k}=
\frac{\om_k-\ep_k}{2\ep_k}\; ,
$$
and $\ep_k$ is the Bogolubov spectrum
\be
\label{37}
\ep_k = \sqrt{\om_k^2 - \Dlt_k^2} \; .
\ee
The condensate separation through the Bogolubov shift (32) is meaningful
only when the particle spectrum (37) touches zero at $k=0$, which gives
\be
\label{38}
\mu=\rho\Phi_0 \; .
\ee
Thus, one comes to the Bogolubov Hamiltonian
\be
\label{39}
H_B =  E_0 + \sum_{k\neq 0} \ep_k b_k^\dgr b_k - \mu N \; ,
\ee
with the ground-state energy
\be
\label{40}
E_0 = \frac{1}{2}\; N\rho\Phi_0 - \; \frac{1}{2}\; \sum_{k\neq 0}
(\om_k - \ep_k) \; .
\ee
Using the chemical potential (38), for the spectrum (35) one has
\be
\label{41}
\om_k =\frac{k^2}{2m} + \rho \Phi_k \; .
\ee
With the diagonal Bogolubov Hamiltonian (39), it is easy to find the
normal,
\be
\label{42}
n_k\equiv\; <a_k^\dgr a_k> \; ,
\ee
and anomalous,
\be
\label{43}
\sgm_k\equiv <a_k a_{-k}>\; ,
\ee
averages. We have
\be
\label{44}
n_k = \frac{\om_k}{2\ep_k} (1 +2\pi_k) - \; \frac{1}{2}
\ee
and
\be
\label{45}
\sgm_k = -\; \frac{\Dlt_k}{2\ep_k} (1 + 2\pi_k) \; ,
\ee
where
\be
\label{46}
\pi_k \equiv\; < b_k^\dgr b_k> \; = \left ( e^{\bt\ep_k} -1
\right )^{-1} \; .
\ee

Now let us turn to investigating the number-of-particle fluctuations. In
the Bogolubov approximation, the number-of-particle operators for condensed,
$\hat N_0$, and noncondensed, $\hat N_1$, particles are uncorrelated, so
that
\be
\label{47}
<\hat N_0 \hat N_1> \; = \; <N_0><\hat N_1>\; .
\ee
Hence, their covariance
$$
{\rm cov}(\hat N_0,\hat N_1) = 0\; .
$$
Therefore
\be
\label{48}
\Dlt^2(\hat N) =\Dlt^2(\hat N_0) + \Dlt^2(\hat N_1) \; .
\ee

Calculating the dispersion $\Dlt^2(\hat N_1)$ for the number-of-particle
operator of noncondensed particles
$$
\hat N_1 =\sum_{k\neq 0} a_k^\dgr a_k \; ,
$$
one has to work out the four-operator expression $<a_k^\dgr a_k a_q^\dgr a_q>$
or, after involving the Bogolubov canonical transformation, one needs to treat
the four-operator terms $<b_k^\dgr b_k b_q^\dgr b_q>$. Such four-operator
products are reorganized by means of the Wick decoupling, which yields
\be
\label{49}
\Dlt^2(\hat N_1)  =\sum_{k\neq 0} \left \{ \left ( 1 +
\frac{2m^2 c_k^4}{\ep_k^2}\right ) \pi_k (1 +\pi_k) +
\frac{m^2 c_k^4}{2\ep_k^2}\right \} \; .
\ee
Here the notation
$$
c_k \equiv \sqrt{\frac{\rho\Phi_k}{m} }
$$
for the effective sound velocity is used, which enters the Bogolubov spectrum
(37) as
\be
\label{50}
\ep_k = \sqrt{(c_k k)^2 +\left ( \frac{k^2}{2m}\right )^2} \; .
\ee

Replacing in Eq. (49) the summation by integration, one gets an infrared
divergence of the type $N\int dk/k^2$. Limiting here the integration by
minimal $k_{min}=1/L$, with $L\propto N^{1/3}$, one gets $\Dlt^2(\hat N_1)
\propto N^{4/3}$, which is anomalous. Remaining in the frame of the discrete
wave vectors ${\bf k}$ does not save the situation, and the dispersion
$\Dlt^2(\hat N_1)$ stands anomalous. But, as follows from the theorem of Sec.
III, the anomalous partial dispersion yields the anomalous total dispersion
$\Dlt^2(\hat N)$, which in the present case is evident from Eq. (48). As
a result, the compressibility (3) diverges as $\kappa_T\propto N^{1/3}$,
which implies the instability of the system as a whole. Thus one would come
to the strange conclusion that stable Bose-condensed systems do not exist.

However, the conclusion on the appearance of anomalous fluctuations in Bose
systems, derived from Eq. (49), is not correct. The mistake here is in the
following. A basic point of the Bogolubov theory is the contraction of the
total Hamiltonian (31) to the quadratic form (34), omitting all terms of the
order higher than two with respect to the operators $a_k$ of noncondensed
particles. The Bogolubov theory is a {\it second-order} theory with respect
to $a_k$. Being in the frame of a second-order theory imposes the restriction
of keeping only the terms of up to the second order when calculating any
physical quantities, and omitting all higher order terms. In working out
the dispersion $\Dlt^2(\hat N_1)$, one meets the fourth-order terms with
respect to $a_k$. Such fourth-order terms are not defined in the second-order
approximation. The calculation of the fourth-order expressions in the
second-order approximation is not self-consistent, i.e., it is incorrect.

A correct calculation of $\Dlt^2(\hat N)$ in the frame of the Bogolubov
theory can be accomplished in the following way. By invoking the relations
(3), (9), and (10), we have
\be
\label{51}
\Dlt^2(\hat N) = N \left \{ 1 + \rho \int [g(\br)-1 ]\; d\br\right \} \; .
\ee
The pair correlation function is
\be
\label{52}
g(\br_{12}) = \frac{1}{\rho^2}
<\psi^\dgr(\br_1)\psi^\dgr(\br_2)\psi(\br_2)\psi(\br_1)>\; ,
\ee
where $\br_{12}=\br_1-\br_2$.

For the field operators, one assumes the Bogolubov shift (32), which taking
into account that in the thermodynamic limit the condensate operator
$\psi_0$ becomes a classical number, can be written as
\be
\label{53}
\psi(\br) = \eta + \psi_1(\br) \; ,
\ee
where the first term is the Bogolubov order parameter
\be
\label{54}
\eta = \; <\psi(\br)>\; = \; <\psi_0> \; ,
\ee
which can be set as $\eta=\sqrt{\rho_0}$, with $\rho_0\equiv N_0/V$. Here
$\eta$ does not depend on $\br$ for a uniform system under consideration.

The pair correlation function (52) can be simplified by invoking the Wick
decoupling. This, however, must be handled with care. A delicate point is
that the Wick decoupling and the Bogolubov shift (53) do not commute with
each other. In the present context, the Wick decoupling is equivalent to
the Hartree-Fock-Bogolubov approximation. The latter does not commute with
the Bogolubov shift. Thus, accomplishing, first, the Bogolubov shift in the
pair correlation function (52), and then using the Hartree-Fock-Bogolubov
approximation for the operators $\psi_1(\br)$, or, what is the same, the
Wick decoupling for the operators $a_k$, with $k\neq 0$, we obtain
\be
\label{55}
g(\br_{12}) = 1 + \frac{2\rho_0}{\rho^2}\; {\rm Re}\left [
\rho_1(\br_1,\br_2) + \sgm_1(\br_1,\br_2)\right ] + \frac{1}{\rho^2}
\left [ |\rho_1(\br_1,\br_2)|^2 +  |\sgm_1(\br_1,\br_2)|^2 \right ] \; .
\ee
Here the Hartree-Fock-Bogolubov approximation for $\psi_1(\br)$ is employed,
resulting in
$$
<\psi_1^\dgr(\br_1)\psi_1(\br_1)\psi_1(\br_2)>\; = 0\; ,
$$
because of the condition $<\psi_1(\br)>=0$, and in
$$
<\psi^\dgr(\br_1)\psi_1^\dgr(\br_2)\psi_1(\br_2)\psi_1(\br_1)>\; =
\rho_1^2 + |\rho_1(\br_1,\br_2)|^2 +  |\sgm_1(\br_1,\br_2)|^2  \; .
$$
The notation is used for the normal average
\be
\label{56}
\rho_1(\br_1,\br_2) \equiv\; < \psi_1^\dgr(\br_2)\psi_1(\br_1)>
\ee
and for the anomalous average
\be
\label{57}
\sgm_1(\br_1,\br_2) \equiv  < \psi_1(\br_2)\psi_1(\br_1)>
\ee
in the real space. These averages are related, by means of the Fourier
transforms
$$
\rho_1(\br_1,\br_2) = \int n_k e^{i{\bf k}\cdot\br_{12}}\;
\frac{d{\bf k}}{(2\pi)^3} \; , \qquad
\sgm_1(\br_1,\br_2) = \int \sgm_k e^{i{\bf k}\cdot\br_{12}}\;
\frac{d{\bf k}}{(2\pi)^3} \; ,
$$
with the normal and anomalous averages (42) and (43), respectively, in the
momentum space.

Note that function (55) possesses the correct limiting behaviour
$$
\lim_{\br_{12}\ra\infty} g(\br_{12}) = 1 \; .
$$
But, if one, first, would make the Hartree-Fock-Bogolubov approximation for
the operators $\psi(\br)$ and, after this, would substitute the Bogolubov
shift (53), then one would get another correlation function with a wrong
limiting behaviour, as is explained in the Appendix A. This is because the
usage of the Wick decoupling, and Hartree-Fock-Bogolubov approximation, for
the operators, represented as sums of several terms, is correct if and only
if all terms in the sum possess the same commutation relations. However, in
the Bogolubov shift (53), the field operators $\psi(\br)$ and $\psi_1(\br)$
do have the same Bose commutation relations, but the term $\eta$ does not 
enjoy such relations. Consequently, the proper way of action is to realize, 
first, the Bogolubov shift (53) and only after this to invoke the
Hartree-Fock-Bogolubov approximation for the operators $\psi_1(\br)$. The
inverse order, as is explained in the Appendix A, is not correct.

For the pair correlation function (55), we find
$$
\int [ g(\br) -1 ]\; d\br = \frac{2\rho_0}{\rho^2}\; \lim_{k\ra 0}
(n_k +\sgm_k) + \frac{1}{\rho^2} \; \int \left ( n_k^2 + \sgm_k^2
\right ) \; \frac{d{\bf k}}{(2\pi)^3} \; .
$$
In the frame of the Bogolubov theory, we have to set $\rho_0=\rho$ and to
omit the terms of the order higher than two with respect to the operators
$a_k$ of noncondensed particles. This means that the terms $n_k^2$ and
$\sgm_k^2$ are to be omitted. Therefore, the number-of-particle dispersion
(51) in the Bogolubov theory is
\be
\label{58}
\Dlt^2(\hat N) =  N\left [ 1 + 2\lim_{k\ra 0} (n_k+\sgm_k)\right ] \; .
\ee
Employing Eqs. (44) to (46), we get
$$
\lim_{k\ra 0} (n_k+\sgm_k) = \frac{1}{2}\left ( \frac{T}{mc^2} - 1
\right ) \; ,
$$
where
$$
c\equiv \lim_{k\ra 0} c_k = \sqrt{\frac{\rho\Phi_0}{m}} \; ,
$$
with
$$
\Phi_0 \equiv \lim_{k\ra 0} \Phi_k = \int \Phi(\br)\; d\br \; .
$$
Then dispersion (58) becomes
\be
\label{59}
\Dlt^2(\hat N) = \frac{T}{mc^2}\; N \; ,
\ee
which is, of course, normal, as it should be for a stable system.
Respectively, the isothermal compressibility
\be
\label{60}
\kappa_T = \frac{\Dlt^2(\hat N)}{\rho TN} = \frac{1}{\rho mc^2}
\ee
is finite.

According to the theorem of Sec. III, if the total dispersion (59) is normal,
then both dispersions of the number-of-particle operators for condensed,
$\Dlt^2(\hat N_0)$, as well as for noncondensed, $\Dlt^2(\hat N_1)$, particles
must be normal. Anomalous fluctuations can arise solely as a result of wrong
calculations, when, e.g., one considers the fourth-order terms $n_k^2$ and
$\sgm_k^2$ in the second-order Bogolubov theory.

\section{Systems with continuous symmetry}

It is easy to show that the same fictitious anomalous fluctuations appear,
not only for Bose systems, but for arbitrary systems, when one treats the
Hamiltonian in the second-order approximation, but intends to calculate
fourth-order expressions. This immediately follows from the analysis of
susceptibilities for arbitrary systems with continuous symmetry, as has been
done by Patashinsky and Pokrovsky [30].  Following Ref. [230], one may consider
an operator $\hat A=\hat A(\vp)$, which is a functional of a field $\vp$. Let
this operator be represented as a sum $\hat A=\hat A_0 +\hat A_1$, where the
first term is quadratic in the field $\vp$, so that $\hat A_0\propto\vp^\dgr
\vp$, while the second term depends on the field fluctuations $\dlt\vp$ as
$\hat A_1\propto\dlt\vp^\dgr\dlt\vp$. Let the system Hamiltonian be taken in
the hydrodynamic approximation, where only the terms quadratic in the field
fluctuations $\dlt\vp$ are retained. The dispersion $\Dlt^2(\hat A)\propto
N\chi$ is proportional to a longitudinal susceptibility $\chi$. The latter
is given by the integral $\int C(\br)d\br$ over the correlation function
$C(\br)\equiv g(\br)-1$, with $g(\br)$ being the pair correlation function.
Calculating $\Dlt^2(\hat A)$, one meets the fourth-order term $<\dlt\vp^\dgr
\dlt\vp\dlt\vp^\dgr\dlt\vp>$. For the quadratic hydrodynamic Hamiltonian,
such fourth-order terms are decoupled by resorting to the Wick theorem. Then
one finds
\be
\label{61}
C(\br) \propto \frac{1}{r^{2(d-2)}}
\ee
for any dimensionality $d>2$. Consequently,
$$
\chi \propto \int C(\br)\; d\br \propto N^{(d-2)/3}
$$
for $2<d<4$. Hence the dispersion is
\be
\label{62}
\Dlt^2(\hat A) \propto N\chi \propto N^{(d+1)/3} \; .
\ee
For $d=3$, this gives $\Dlt^2(\hat A)\propto N^{4/3}$, that is, the same
anomalous dispersion as $\Dlt^2(\hat N)$ for Bose systems. But this implies
that the related susceptibility diverges as $\chi\propto N^{1/3}$, which
tells that the considered system is unstable. If this would be correct, it
would mean that there are no stable systems with continuous symmetry. For
instance, there could not exist magnetic systems, described by the Heisenberg
Hamiltonian. Liquid helium also could not exist as a stable system.

The existence or absence of anomalous fluctuations does not depend on the
statistical ensemble used. Thus, in the frame of the same calculational
procedure, the particle fluctuations are the same, being either anomalous
or normal, depending on the chosen procedure, for all ensembles, whether
canonical, grand canonical, or microcanonical [31].

It is worth emphasizing that such fictitious anomalous fluctuations arise
not just at a phase transition point, which would not be surprising, but
everywhere below this point, in the whole region of existence of the
considered system. That is, everywhere below the phase transition points
such systems would not be stable. As is evident, such a strange conclusion
is physically unreasonable. Fortunately, the explanation for the occurrence
of anomalous fluctuations is rather simple: They arise solely due to an
incorrect calculational procedure, when the fourth-order terms are treated
by a second-order theory, such as the hydrodynamic approximation. No anomalous
fluctuations happen, if all calculations are done self-consistently, being
defined in the frame of the given approximation. 

Another popular way of incorrectly obtaining thermodynamically anomalous 
particle fluctuations for systems with continuous symmetry is as follows. 
One uses the representation
\be
\label{63}
\psi(\br) = e^{i\hat\vp(\br)}\; \sqrt{\hat n(\br)}
\ee
for the field operator, in which $\hat n(\br)\equiv\psi^\dgr(\br)\psi(\br)$ 
is the operator of particle density and $\hat\vp(\br)$ is the phase operator. 
The latter is assumed to be Hermitian in order to preserve the correct 
definition of the density operator,
$$
\psi^\dgr(\br)\psi(\br) =\sqrt{\hat n(\br)}\; 
e^{-i\hat\vp^+(\br)+i\hat\vp(\br)} \sqrt{\hat n(\br)} = 
\hat n(\br) \; .
$$
It is easy to show that from the representation (63) it follows that the 
density and phase operators are canonically conjugated, satisfying the 
commutation relation
$$
\left [ \hat n(\br),\; \hat\vp(\br')\right ] = i\dlt(\br-\br') \; .
$$
For the first-order correlation function, one has
$$
<\psi^\dgr(\br)\psi(0)>\; = \; <\sqrt{\hat n(\br)\hat n(0)}\;
\exp\left\{ - i\left [\hat\vp(\br) - \hat\vp(0)\right ] \right \} > \; .
$$
Then one assumes that the temperature is asymptotically low, $T\ra 0$, 
such that there are no density fluctuations, and one can replace the 
operator $\hat n(\br)$ by its average $\rho(\br)\equiv<\hat n(\br)>$. 
This is equivalent to the usage, instead of the representation (63), of 
the representation 
\be
\label{64}
\psi(\br) = \sqrt{\rho(\br)}\; e^{i\hat\vp(\br)} \; .
\ee
One also supposes that the phase fluctuations are very small, so that 
one can employ the following averaging:
\be
\label{65}
<\exp\left\{ -i\left [ \hat\vp(\br) -\hat\vp(0)\right ] \right\} >\;
= \; \exp\left\{ -\; \frac{1}{2} <\left [ \hat\vp(\br) -\hat\vp(0)
\right ]^2 > \right \} \; .
\ee
As a result, the first-order correlation function reduces to
$$
<\psi^\dgr(\br)\psi(0)> \; = \rho(\br) \exp\left\{ -\; \frac{1}{2} 
<\left [ \hat\vp(\br) -\hat\vp(0) \right ]^2 > \right \} \; .
$$
Treating $\hat\vp(\br)$ as a small quantity, one also expands the 
exponentials in powers of $\hat\vp(\br)$. Similarly, one treats the 
second-order correlation functions. Finally, one comes to the same 
expressions as in Eqs. (61) and (62), with the thermodynamically 
anomalous fluctuations, $\Dlt^2(\hat N_1)\propto N^{4/3}$, for the 
three-dimensional space.

The main mistake in such calculations is the same as has been made 
above. All calculations have been based on the assumption that both 
the density a and phase fluctuations are rather weak, so that the 
hydrodynamic approximation could be invoked. The latter implies that 
all statistical averages are treated in the hydrodynamic approximation, 
with a Hamiltonian quadratic in the operators. For instance, it is well 
known [32] that Eq. (65) is valid solely for quadratic Hamiltonians. 
For finding $\Dlt^2(\hat N_1)$, one needs to consider the fourth-order 
terms in phase operators. Of course, there is no sense in calculating 
the forth-order terms in the frame of a second-order theory, such as 
the hydrodynamic approximation.

Moreover, the representations (63) and (64), as such, are principally 
incorrect. This is shown in the Appendix B. A correct definition of the 
phase operator requires a much more elaborate technique, as can be 
inferred from the review articles [33--36]. Since the representations 
(63) and (64), actually, do not exist, all conclusions derived on their 
basis, even involving no further approximations, are not reliable.

\section{Breaking of Gauge Symmetry}

In Section IV, considering the ideal uniform Bose gas, we found that 
its particle fluctuations are thermodynamically anomalous, with the 
corresponding dispersion $\Dlt^2(\hat N)\propto N^2$. This anomaly is 
due to the condensate fluctuations, since $\Dlt^2(\hat N_0)\propto N^2$. 
Really, for an ideal uniform gas, one has
\be
\label{66}
\Dlt^2(\hat N) = \sum_k n_k (1+n_k) \; .
\ee
From here, separating the terms with $\bk=0$ and $\bk\neq 0$, we get
$$
\Dlt^2(\hat N_0) = N_0 (1 + N_0) \; , \qquad
\Dlt^2(\hat N_1) = \sum_{k\neq 0} n_k(1+n_k) \; .
$$
Since $N_0\propto N$, we find $\Dlt^2(\hat N_0) \propto N^2$.

The situation can be made even more dramatic by generalizing it to 
the case of interacting particles. To this end, let us consider an 
interacting system that can be treated by perturbation theory starting 
with a mean-field approximation, such as the Hartree-Fock approximation. 
In the frame of the latter, the particle dispersion can be shown [37] to 
have the same form as in Eq. (66). Then, irrespectively of the concrete 
expression for the momentum distribution of particles $n_k$, the global 
dispersion $\Dlt^2(\hat N)$ will be thermodynamically anomalous because 
of the anomalous term $\Dlt^2(\hat N_0)\propto N^2$. Hence, one could 
conclude that all systems with the Bose-Einstein condensate would be 
unstable.

One often states that the appearance of this anomaly is the defect of the 
grand canonical ensemble. However, this is not correct. As is mentioned in 
Section IV, the anomalous condensate fluctuations are fictitious and can be 
removed by breaking the gauge symmetry.

Hohenberg and Martin [38] noticed that the appearance of such fictitious 
divergences is a common feature of theories possessing gauge symmetry, 
but breaking the latter would eliminate the divergences resulting from the 
condensate fluctuations. Ter Haar [25] showed explicitly how the anomalous 
condensate fluctuations can be removed after breaking the gauge symmetry 
for an ideal uniform Bose gas. In the present section, we demonstrate that, 
in general, the gauge-symmetry breaking eliminates the anomalous condensate 
fluctuations for arbitrary systems, whether interacting or not.

A known method for lifting a system symmetry of any nature is the method 
of infinitesimal sources, introduced by Bogolubov [29,39]. There are also 
several other methods of symmetry breaking, as is reviewed in Ref. [40]. In 
the case of gauge symmetry, one has to be cautious by chosing the way of 
its breaking. The standard method of infinitesimal sources may not always 
lead to the desired symmetry breaking, as is shown by a counterexample in 
the Appendix C.

To break the gauge symmetry in a Bose system, one has to resort to the 
Bogolubov shift [29,39]. The latter, keeping in mind the most general 
statistical system, whether equilibrium or nonequilibrium, uniform or 
nonuniform, writes as
\be
\label{67}
\psi(\br,t) = \eta(\br,t) + \psi_1(\br,t) \; ,
\ee
where $t$ is time. The first term here is the condensate wave function, 
assumed to be not identically zero in the presense of the Bose-Einstein 
condensate. The second term in Eq. (67) is the field operator of 
noncondensed particles, satisfying the same Bose commutation relations 
as $\psi(\br,t)$. The correct separation of condensed and noncondensed 
particles presupposes the orthogonality condition
\be
\label{68}
\int \eta^*(\br,t)\psi_1(\br,t)\; d\br = 0 \; ,
\ee
which exculdes the double counting of the degrees of freedom. In what 
follows, just for brevity, we shall write $\psi(\br)$ instead of 
$\psi(\br,t)$, understanding that, generally, the time variable $t$ 
does enter the dependence of the field operator, $\psi(\br)=\psi(\br,t)$.

For the theory of Bose systems, it is extremely important to specify the 
spaces of states, which the field operators are defined on. Thus, the field 
operators $\psi(\br)$ and $\psi^\dgr(\br)$ are defined on the Fock space
${\cal F}(\psi)$ generated by the operator $\psi^\dgr(\br)$. This means the 
following [41]. There exists a vacuum state $|0>$, for which
\be
\label{69}
\psi(\br)|0>\; = \; 0 \; .
\ee
The Fock space ${\cal F}(\psi)$ is the space of all states
$$
\vp = \sum_{n=0}^\infty \; \frac{1}{\sqrt{n!}} \; \int
f_n(\br_1,\ldots,\br_n)\; \prod_{i=1}^n \psi^\dgr(\br_1) 
\; d\br_i | 0> \; ,
$$
in which $f_n(\br_1,\ldots,\br_n)$ is a square-integrable function 
symmetric with respect to the permutation of any pair of its variables.

It is easy to notice that the state $|0>$, which is a vacuum state for 
$\psi(\br)$, is not a vacuum for $\psi_1(\br)$, since
$$
\psi_1(\br)|0> \; = \; -\eta(\br)|0> \; \not\equiv \; 0 \; .
$$
Consequently, there should exist another state $|0>_1$ satisfying the 
condition
\be
\label{70}
\psi_1(\br)|0>_1 = 0 \; ,
\ee
being a vacuum for $\psi_1(\br)$. In turn, the state $|0>_1$, which is 
a vacuum for $\psi_1(\br)$, is not a vacuum for $\psi(\br)$, as far as
$$
\psi(\br)|0>_1 = \eta(\br)|0>_1 \not\equiv 0 \; .
$$

The Bogolubov shift (67) is a particular case of canonical transformations 
[42]. The operators $\psi(\br)$ and $\psi_1(\br)$ can be connected with 
each other by means of the transformation
\be
\label{71}
\hat C \equiv \exp \left \{ \int \left [ \eta^*(\br)\psi(\br) -
\eta(\br)\psi^\dgr(\br)\right ] d\br \right \} 
\ee
and its inverse
\be
\label{72}
\hat C^{-1} = \exp \left \{ - \int \left [ \eta^*(\br)\psi(\br) -
\eta(\br)\psi^\dgr(\br)\right ] d\br \right \} \; .
\ee
Using these transformations, one has
\be
\label{73}
\psi(\br) = \hat C \psi_1(\br) \hat C^{-1}
\ee
and
\be
\label{74}
\psi_1(\br) = \hat C^{-1} \psi(\br) \hat C \; .
\ee
Then it becomes clear that the vacuum for $\psi_1(\br)$ is
\be
\label{75}
|0>_1 \; = \hat C^{-1}|0> \; .
\ee

The vacua $|0>$ and $|0>_1$ are mutually orthogonal. This can be shown by 
employing the Baker-Hausdorff formula, which for two operators $\hat A$ and 
$\hat B$, whose commutator $[\hat A,\hat B]$ is proportional to the unity 
operator, reads as
$$
e^{\hat A+\hat B} = e^{\hat A} e^{\hat B} \exp \left ( -\; \frac{1}{2}
\left [\hat A,\; \hat B\right ]\right ) \; .
$$
Using this for transformation (72), we have
\be
\label{76}
\hat C^{-1} = \exp\left\{ \int \eta(\br) \psi^\dgr(\br) \; d\br\right \}\;
\exp\left\{ - \int \eta^*(\br) \psi(\br) \; d\br\right \}\;
\exp\left\{ -\frac{1}{2}\;  \int |\eta(\br)|^2 \; d\br\right \} \; .
\ee
Acting on the vacuum $|0>$, we find
\be
\label{77}
\hat C^{-1}|0> \; = \exp\left\{ -\frac{1}{2}\;  \int |\eta(\br)|^2 \; 
d\br\right \} \; \exp\left\{ \int \eta(\br) \psi^\dgr(\br) \; d\br
\right \} |0> \; .
\ee
This is nothing but the coherent state [43], being the eigenstate of the 
destruction operator,
\be
\label{78}
\psi(\br)|\eta> \; = \; \eta(\br)|\eta> \; ,
\ee
and having in the coordinate representation [44] the form
\be
\label{79}
|\eta> \; = \; \eta_0 \exp\left\{ \int \eta(\br)\psi^\dgr(\br)\;
d\br \right \} | 0> \; ,
\ee
with the normalization factor
$$
|\eta_0| = \exp\left\{ -\frac{1}{2}\;  \int |\eta(\br)|^2 \; 
d\br\right \} \; .
$$
Respectively, the condensate wave function 
$$
\eta(\br) = \; <\eta|\psi(\br)|\eta >
$$
is nothing but the coherent field related to the coherent state $|\eta>$.

In this way, the vacuum (75) is the coherent state (79),
\be
\label{80}
|0>_1 \; = \hat C^{-1}|0> \; = \; |\eta>\; .
\ee
The scalar product of the vacua $|0>$ and $|0>_1$ is
\be
\label{81}
<0|0>_1 \; = \; <0|\eta>\; = \; \exp\left\{ -\frac{1}{2}\;  
\int |\eta(\br)|^2 \; d\br\right \} \; .
\ee
By its definition, the condensate wave function gives the condensate 
density
\be
\label{82}
\rho_0(\br) \equiv |\eta(\br)|^2 \; .
\ee
The number of condensed particles
\be
\label{83}
N_0 = \int \rho_0(\br)\; d\br \; ,
\ee
in the presence of the condensate, is not zero, but is macroscopic in the 
sense that $N_0\propto N\ra\infty$. Therefore the scalar product
\be
\label{84}
<0|0>_1 \; = \exp \left ( -\; \frac{1}{2}\; N_0 \right )
\ee
becomes zero in the thermodynamic limit,
\be
\label{85}
<0|0>_1 \; \simeq \; 0 \qquad (N\ra \infty) \; .
\ee
This tells that the vacua $|0>$ and $|0>_1$ are asymptotically orthogonal. 
The Fock spaces ${\cal F}(\psi)$ and ${\cal F}(\psi_1)$, generated from the 
related vacua, are orthogonal to each other, except just the sole state 
$|0>_1=|\eta>$, which is the vacuum for ${\cal F}(\psi_1)$ and the coherent 
state, defined by Eq. (78), in ${\cal F}(\psi)$. However, having the sole 
common state for two infinite-dimensional spaces means the intersection 
of zero measure. Moreover, the influence of this intersection is eliminated 
by means of the orthogonality condition (68).

Thus, there are two different vacua $|0>$ and $|0>_1$ and two mutually
orthogonal Fock spaces ${\cal F}(\psi)$ and ${\cal F}(\psi_1)$, generated 
by the field operators $\psi^\dgr$ and $\psi_1^\dgr$, respectively. The 
operator (71) transforms ${\cal F}(\psi_1)$ into ${\cal F}(\psi)$, while 
the operator (72) transforms ${\cal F}(\psi)$ into ${\cal F}(\psi_1)$. 
There is no self-adjoint operator $\hat C^+$ that would be defined on 
the same space as $\hat C$. Therefore the operator $\hat C$ is nonunitary 
and the transformations (73) and (74) cannot be treated as unitary. The 
field operators $\psi$ and $\psi_1$ are defined on different spaces. One 
says that such operators realize unitary nonequivalent operator 
representations of canonical commutation relations [45].

Breaking the gauge symmetry by the Bogolubov shift (67), one, actually, 
passes from the Fock space ${\cal F}(\psi)$ to the space ${\cal F}(\psi_1)$. 
Since the left-hand and right-hand sides of Eq. (67) are defined on different 
spaces, this equation should be understood as a transformation
$$
\psi(\br) \; \longrightarrow \; \eta(\br) +\psi_1(\br) \; .
$$
Separating the zero-momentum mode for a uniform Bose gas, with replacing 
this term by a nonoperator quantity,
$$
\psi_0 = \frac{a_0}{\sqrt{V}} \; \ra \; \sqrt{\rho_0} \; ,
$$
as has been done in Section V, is mathematically equivalent to the 
Bogolubov shift [46]. The representation of the operators of observables, 
expressed through the field operators $\psi_1$, and defined on the Fock 
space ${\cal F}(\psi_1)$, can be called the Bogolubov representation.

In the Bogolubov representation, the operator of condensed particles, 
according to Eqs. (82) and (83), is a nonoperator quantity, $\hat N_0=N_0$. 
Hence, the dispersion of the latter is zero, $\Dlt^2(N_0)=0$. Consequently, 
the dispersion of the total number-of-particle operator
$$
\Dlt^2(\hat N) = \Dlt^2(\hat N_1)
$$
is completely defined by the dispersion of the operator $\hat N_1$ of 
noncondensed particles. Thus, the anomalous $N^2$ dispersion of the 
condensate particles is removed in the Bogolubov representation.

Considering the ideal uniform Bose gas of Section IV in the Bogolubov 
representation, we do not meet the $N^2$-anomalous condensate fluctuations. 
Nevertheless, particle fluctuations, characterized by the dispersion 
$\Dlt^2(\hat N_1)\propto N^{4/3}$, remain thermodynamically anomalous. 
That is, this gas, anyway, is unstable. This conclusion does not depend 
on whether the grand canonical or canonical ensemble has been used. Of 
course, in the latter, where the total number of particles is fixed, the 
related dispersion is not defined. However, one can calculate the 
compressibility
$$
\kappa_T = -\; \frac{1}{V} \left ( \frac{\prt P}{\prt V}\right )^{-1}_{TN}
= \frac{1}{V} \left ( \frac{\prt^2 F}{\prt V^2}\right )^{-1}_{TN} \; ,
$$
where $F$ is free energy. For the ideal uniform Bose gas below $T_c$, 
one has [2] $\prt P/\prt V=0$, hence, $\kappa_T\ra\infty$, which implies 
instability. The latter is an intrinsic feature of the uniform ideal Bose 
gas [13]. Including particle interactions stabilizes the gas, as is shown 
in Section V. The ideal Bose gas can also be stabilized by trapping it in 
an external confining potential, such as the harmonic potential [47,48], 
though not all power-law potentials are able to stabilize the system [49].

The message of this section is that accurately defining the symmetry 
properties of the given system helps to avoid the appearance of unphysical 
instabilities. Although there also exist systems, such as the ideal uniform 
Bose-condensed gas, which are intrinsically unstable.

\section{Notion of Representative Ensembles}

The consideration of the previous Section VII demonstrates the importance 
of accurately defining the system under investigation. It is not sufficient 
to chose a statistical ensemble, but often it is also necessary to formulate 
additional conditions specifying the features of the given system, thus, 
avoiding the appearance of spurious instabilities. For instance, one can take 
the grand canonical ensemble without breaking the gauge symmetry or one may 
employ the grand canonical ensemble with the gauge symmetry breaking. This 
means that, in general, there may exist not just the sole grand canonical 
ensemble or the sole canonical one, but there can exist several such 
ensembles. This problem of the ensemble nonuniqueness is just another way 
of formulating the problem of the nonuniqueness of the Fock space and of 
the existence of unitary nonequivalent operator representations, which is 
explained in the previous Section VII.

Thus, for the correct description of a physical system, it is necessary 
to equip the chosen statistical ensemble by additional conditions required 
for accurately taking account of the system features. Only such an equipped 
ensemble will correctly represent the considered system, that is, will be 
a {\it representative ensemble}.

The idea of the representative ensembles goes back to Gibbs himself 
[50], who mentioned the necessity of taking into account all additional 
information known about the considered system, such as the system symmetry, 
the existence of integrals of motion, and so on. The importance of employing 
representative ensembles for an adequate description of statistical systems 
was emphasized by ter Haar [51,52]. A detailed discussion of mathematical 
techniques, required for the correct definition of representative ensembles, 
can be found in the review papers [40,53]. In the language of reduced 
density matrices, the latter have to satisfy specific constraints in order 
to correctly represent a given statistical system [54].

Systems, exhibiting Bose-Einstein condensation, serve as a very good 
example demonstrating the importance of taking into account their specific 
features in order to correctly describe their behaviour. Rich properties of 
these systems require to be very attentive in formulating the corresponding 
representative ensemble. Forgetting to impose the appropriate constraints, 
specifying the system properties, may lead to self-inconsistent calculations 
and the appearance of spurious instabilities. In Section V, the example was 
given of a weakly-interacting equilibrium uniform Bose gas. Now we shall 
formulate a general approach to Bose systems with arbitrarily strong 
interactions, being, in general, nonuniform and not necessarily equilibrium. 
We shall stress the constraints that are compulsory for defining a 
self-consistent theory, which, for equilibrium systems, results in a 
representative ensemble, free of fictitious instabilities.

First of all, as is explained in Section VII, we have to break the gauge 
symmetry by means of the Bogolubov shift, replacing the field operator 
$\psi(\br,t)$, acting in the Fock space ${\cal F}(\psi)$, by the operator
\be
\label{86}
\tilde\psi(\br,t) \equiv \eta(\br,t) + \psi_1(\br,t) \; ,
\ee
defined on the Fock space ${\cal F}(\psi_1)$. In what follows, we shall 
again omit the time variable in order to simplify the notation. The first 
term in the right-hand side of Eq. (86) is the condensate wave function 
and the second term is the field operator of noncondensed particles. The 
replacement $\psi(\br)\ra\tilde\psi(\br)$ yields to the passage from 
the operator representation on the Fock space ${\cal F}(\psi)$ to the 
unitary nonequivalent operator representation, the Bogolubov representation,
on the space ${\cal F}(\psi_1)$ only if the condensate wave function 
$\eta(\br,t)$ is not identically zero.

The energy operator has now to be expressed through the field operators 
(86), which yields the Hamiltonian
$$
\hat H = \int \tilde\psi^\dgr(\br) \left (  -\; \frac{\nabla^2}{2m} + 
U \right ) \tilde\psi(\br) \; d\br +
$$
\be
\label{87}
+ \frac{1}{2} \; \int \tilde\psi^\dgr(\br) \tilde\psi^\dgr(\br')
\Phi(\br-\br') \tilde\psi(\br') \tilde\psi(\br)\; d\br d\br'\; ,
\ee
in which $U=U(\br,t)$ is an external field. The corresponding Lagrangian 
is
\be
\label{88}
\hat L \equiv \int \tilde\psi^\dgr(\br) i\; \frac{\prt}{\prt t} \;
\tilde\psi(\br) \; d\br \; - \; \hat H \; .
\ee

It is important to stress that, contrary to a system without condensate, 
where there is just one field operator variable $\psi$, in a Bose-condensed 
system, there appear two variables $\eta$ and $\psi_1$, or one can take as 
two variables $\eta$ and $\tilde\psi$. The condensate wave function defines 
the condensate density (82). The operator of the total number of particles
\be
\label{89}
\hat N = \int \tilde\psi^\dgr(\br) \tilde\psi(\br)\; d\br
\ee
is expressed through $\tilde\psi$. Respectively, there are two normalization 
conditions. One condition is for the condensate wave function normalized to 
the number of condensed particles
\be
\label{90}
N_0 = \int |\eta(\br)|^2 \; d\br \; .
\ee
And another normalization condition is for $\tilde\psi$ normalized to 
the total number of particles $N=<\hat N>$, i.e.,
\be
\label{91}
N = \int <\tilde\psi^\dgr(\br)\tilde\psi(\br)> d\br \; .
\ee
Here and everywhere in this section, the angle brackets imply the averaging 
over the Fock space ${\cal F}(\psi_1)$.

Hamiltonian (87), with the field operator (86), contains the terms linear in 
$\psi_1$, because of which the average $<\psi_1>$ may be nonzero. However, a 
nonzero $<\psi_1>$ would, in general, lead to the nonconservation of quantum 
numbers, such as spin and momentum, which would be unphysical. Therefore, it 
is necessary to impose the constraint for the conservation of quantum numbers,
\be
\label{92}
<\psi_1(\br)>\; = \; 0 \; .
\ee
In this way, three conditions are to be valid for a Bose-condensed 
system, two normalization conditions (90) and (91), and the 
quantum-number conservation constraint (92).

The most general procedure of deriving the equations of motion is 
by looking at the extrema of the action, under the given additional 
conditions. In our case, the effective action is
\be
\label{93}
A[\eta,\; \psi_1] = \int \left ( \hat L + \mu_0 N_0 + \mu \hat N  +
\hat\Lambda \right ) \; dt \; .
\ee
Here, $\hat L$ is the Lagrangian (88). The second and third terms in 
the integral (93) preserve the normalization conditions (90) and (91). 
And the role of the term
\be
\label{94}
\hat \Lambda \equiv \int \left [ \lbd(\br)\psi_1^\dgr(\br) +
\lbd^*(\br)\psi_1(\br)\right ]\; d\br
\ee
is to satisfy the quantum-number conservation constraint (92). The 
Lagrange multipliers $\lbd(\br)$ have to be chosen so that to cancel in 
Eq. (87) the terms linear in $\psi_1$. The absence of such linear terms 
in the Hamiltonian, as is known [42], is necessary and sufficient for 
the validity of condition (92). By introducing the effective grand 
Hamiltonian
\be
\label{95}
H[\eta,\; \psi_1] \equiv \hat H - \mu_0N_0 - \mu\hat N -
\hat\Lambda
\ee
and the resulting Lagrangian 
\be
\label{96}
L[\eta,\; \psi_1] = \int \left [ \eta^*(\br)i\; \frac{\prt}{\prt t}\;
\eta(\br) +\psi_1^\dgr(\br) i\; \frac{\prt}{\prt t}\; \psi_1(\br)
\right ] \; d\br - H[\eta,\; \psi_1] \; ,
\ee
the effective action (93) can be rewritten as
\be
\label{97}
A [\eta,\; \psi_1] = \int L [\eta,\; \psi_1] \; dt \; .
\ee
According to the standard prescription, the equations of motion are 
obtained from the variational principle determining the extremum of 
the action functional (97). These variational equations are
\be
\label{98}
\frac{\dlt A[\eta,\psi_1]}{\dlt\eta^*(\br,t)} = 0 \; ,
\ee
where, for generality, the time variable is written explicitly, and
\be
\label{99}
\frac{\dlt A[\eta,\psi_1]}{\dlt\psi_1^\dgr(\br,t)} = 0 \; .
\ee
From Eqs. (95), (96), and (97), it follows that Eqs. (98) and (99) are
identical to the variational equations
\be
\label{100}
i\; \frac{\prt}{\prt t}\; \eta(\br,t) = 
\frac{\dlt H[\eta,\psi_1]}{\dlt\eta^*(\br,t)} \; ,
\ee
with the effective grand Hamiltonian (95), and
\be
\label{101}
i\; \frac{\prt}{\prt t}\; \psi_1(\br,t) = 
\frac{\dlt H[\eta,\psi_1]}{\dlt\psi_1^\dgr(\br,t)} \; .
\ee
Explicitly, Eq. (100) is
$$
i\; \frac{\prt}{\prt t}\; \eta(\br,t) = \left ( -\; 
\frac{\nabla^2}{2m} + U -\ep\right ) \eta(\br ) +
$$
\be
\label{102}
+ \int \Phi(\br-\br') \left [ |\eta(\br')|^2 \eta(\br) +
\hat X(\br,\br') \right ] \; d\br' \; ,
\ee
where $\ep\equiv\mu_0+\mu$ and again, for short, the time dependence 
is omitted. Equation (101)  yields
$$
i\; \frac{\prt}{\prt t}\; \psi_1(\br,t) = \left ( -\; 
\frac{\nabla^2}{2m} + U -\mu \right )  \psi_1(\br) +
$$
\be
\label{103}
+ \int \Phi(\br-\br') \left [ |\eta(\br')|^2 \psi_1(\br) +
\eta^*(\br')\eta(\br)\psi_1(\br') + 
\eta(\br')\eta(\br)\psi_1^\dgr(\br')
+ \hat X(\br,\br')\right ]\; d\br' \; .
\ee
Here the notation
$$
\hat X(\br,\br') \equiv \psi_1^\dgr(\br')\psi_1(\br')\eta(\br)
+ \psi_1^\dgr(\br')\eta(\br')\psi_1(\br) + 
$$
\be
\label{104}
+ \eta^*(\br')\psi_1(\br')\psi_1(\br)
+ \psi_1^\dgr(\br')\psi_1(\br')\psi_1(\br)
\ee
is used. Averaging Eq. (102), we obtain the equation for the 
condensate wave function
$$
i\; \frac{\prt}{\prt t}\; \eta(\br,t) = \left ( -\; 
\frac{\nabla^2}{2m} + U -\ep \right ) \eta(\br) +
$$
\be
\label{105}
+ \int \Phi(\br-\br') \left [ \rho(\br')\eta(\br) +
\rho_1(\br,\br')\eta(\br') +\sgm_1(\br,\br')\eta^*(\br') + 
<\psi_1^\dgr(\br')\psi_1(\br')\psi_1(\br)> \right ]\; d\br' \; ,
\ee
in which the total density of particles
$$
\rho(\br) =\rho_0(\br) +\rho_1(\br)
$$
is the sum of the condensate density (82) and of the density of 
noncondensed particles
$$
\rho_1(\br) \equiv \; <\psi_1^\dgr(\br)\psi_1(\br)>\; ;
$$
also the notation is used for the normal density matrix
$$
\rho_1(\br,\br') \equiv \; < \psi_1^\dgr(\br')\psi_1(\br)> \; ,
$$
and the so-called anomalous density matrix
$$
\sgm_1(\br,\br') \equiv \; <\psi_1(\br')\psi_1(\br)> \; ,
$$
which is nonzero because of the broken gauge symmetry.

It is not our goal to study here particular consequencies of the approach 
sketched above. The sole aim of the example of this section is to illustrate 
the way of constructing a representative ensemble for a rather nontrivial 
system. This is done by accurately specifying the basic system properties, 
such as the broken gauge symmetry, normalization conditions, and the 
quantum-number conservation condition. Following the most general procedure 
of action variation, under the specified conditions, one automatically obtains
an effective Hamiltonian and the related exact equations of motion. It is 
possible to show [37] that the latter guarantee the correct behaviour for the 
spectrum of collective excitations, the validity of all conservation laws, and
the absence of unphysical instabilities.

It may happen in some lower-order approximations that there is no need to 
invoke all of the conditions discussed above. This, for instance, occurs in 
the Bogolubov approximation of Section IV. In this approximation, one assumes 
that $N_0\ra N$, hence $\mu_0\ra 0$. Also, for a uniform gas, the Hamiltonian 
term of the first order in $\psi_1$ vanishes itself, while the terms of the 
third and fourth order in $\psi_1$ are neglected in the Bogolubov second-order
approximation. Because of this, there is no necessity of introducing the term 
(94). However, all these conditions are to be taken into account when going 
to higher-order approximations. In the other case, the defined ensemble may 
occur to be nonrepresentative, which can result in physical inconsistences 
and fictitious instabilities.

Correctly defining a representative ensemble is also crucially important for 
the problem of equivalence of statistical ensembles, which is discussed in 
the next section.

\section{Problem of Ensemble Equivalence}

The examples of the previous sections show that the stability properties 
of a system can be different in different ensembles. More general, the 
same physical quantity may be different, being calculated in two different 
ensembles. Does this mean the failure of the basic principle of statistical 
mechanics, stating the equivalence of ensembles for large systems? This 
question is analyzed in the present section.

First of all, let us stress that, as is clear from the previous sections, 
a physical system and a describing it ensemble do not exist separately, but 
they are intimately connected. A correct formulation of an ensemble does 
presuppose that it includes the information on the main system features. 
An ensemble, which is adequate for the given physical system, is only that, 
which properly represents the system, that is, a {\it representative 
ensemble}. But if there are two representative ensembles for the same system, 
then, by their definition, they must yield identical results for the same 
physical quantities. In the other case, at least one of these ensembles does 
not correctly describe the system, hence, is not representative. Also, in the 
case of equilibrium, it is meaningful to talk only about stable systems, as 
far as an unstable system cannot be in absolute equilibrium. Thus, in terms 
of representative ensembles, the following statement is straightforward: {\it 
Two ensembles are equivalent if and only if both of them are representative 
for the given stable system}. Conversely, when two ensembles are not 
equivalent, then at least one of them is not representative. An ensemble that 
is not representative for the given system may be representative for some 
other system. However, there is no any reason to require that two ensembles 
applied to two different physical systems be equivalent. Ensemble 
nonequivalence, vaguely formulated, is a rather artificial nonphysical 
problem caused by an improper usage of ensembles not representing the 
considered system.

To be more correct, let us recall that, generally, one distinguishes 
two types of ensemble equivalence, thermodynamic and statistical. In 
thermodynamics, a physical system is characterized by thermodynamic 
potentials, each of which is a function of its natural thermodynamic 
variables [1--7]. The system is stable, when thermodynamic potentials 
enjoy the property of convexity or concavity with respect to the appropriate 
variables. The thermodynamic potentials, expressed through different 
thermodynamic variables, are connected with each other by Legendre transforms 
[1--7]. All thermodynamic characteristics are defined as derivatives of 
thermodynamic potentials. When the latter are connected by Legendre transforms
and correspond to a stable (in the sense of the convexity or concavity 
property of the potentials) system, then the thermodynamic characteristics, 
calculated in different ensembles, coincide with each other. Summarizing, 
the concept of thermodynamic equivalence can be formulated as follows:

\vskip 2mm
{\it Thermodynamic equivalence}. Two ensembles , representing a stable 
physical system, are thermodynamically equivalent if and only if their 
thermodynamic potentials are mutually connected by Legendre transforms.

\vskip 2mm
A rigorous proof of this statement for the case of the macrocanonical 
and canonical ensembles can be found in Refs. [55,56]. Several examples 
of systems with long-range interactions have been considered, whose 
microcanonical entropy is not a concave function of energy [55--58]. 
The internal energy of such systems, though being nonadditive, can be 
made extensive by means of the Kac-Uhlenbeck-Hemmer normalization [59] 
yielding a well defined thermodynamic limit. The canonical free energy 
is a concave function of inverse temperature, but the microcanonical 
entropy is not a concave function of energy. This does not allow to 
use the Legendre transform in both directions [55,56]. The nonconcavity 
of the microcanonical entropy results in the appearance, for some range 
of energies, of negative specific heat, while in the canonical ensemble 
specific heat is always positive. Because of this, one tells that, for 
such models with long-range interactions, the microcanonical and 
canonical ensembles are not equivalent. However, a microcanonical 
ensemble with a nonconcave entropy does not represent a stable physical 
system, i.e., this ensemble is not representative. As is explained above, 
there is no sense to compare nonrepresentative ensembles, which are not 
obliged to be equivalent. To make the microcanonical ensemble 
representative, it must be complimented by the concavity construction 
rendering stability again. After this, it becomes representative and 
completely equivalent to the canonical ensemble.

Nonconcave microcanonical entropy and negative specific heat are 
also known for gravitating systems, as is reviewed in Refs. [60,61]. 
To avoid the negative specific heat, one can again invoke a concavity 
construction or to use the canonical ensemble. However, contrary to 
other models with long-range interactions, the energy of gravitating 
systems, being proportional to $N^{5/3}$, cannot be made extensive, 
which does not allow the existence of the thermodynamic limit. For 
gravitating systems, the condition of global equilibrium [62]
\be
\label{106}
\frac{E}{N} \geq \; const \; < \; 0
\ee
is not valid. Therefore, they may be in principle unstable, which 
makes questionable the application for their description of equilibrium 
statistical mechanics.

The notion of statistical equivalence of ensembles is based on the 
comparison of the averages of observable quantities calculated in 
different ensembles. To concretize this, let us consider the operators 
of observables $\hat A$ defined on a Fock space ${\cal F}$. The set of 
all these operators forms the algebra of observables ${\cal A}\equiv
\{\hat A\}$. The statistical state is defined [44,63] as the set 
$<{\cal A}>\equiv\{<\hat A>\}$ composed of all statistical averages 
for the algebra of observables. The calculation of the averages is 
defined in the standard way as the trace of $\hat A$, with a 
statistical operator corresponding to the chosen ensemble. Let us 
define as $<{\cal A}>_\mu$ the statistical state related to the grand 
canonical ensemble, with a chemical potential $\mu$. For short, the 
dependence of the state on temperature $T$ and volume $V$ is not shown 
explicitly. For instance, the average density is
\be
\label{107}
\rho = \frac{N}{V} \; , \qquad N= \; <\hat N>_\mu \; .
\ee

Suppose, we wish to compare the grand canonical and canonical ensembles. 
Recall that the general structure of the Fock space is a direct sum
\be
\label{108}
{\cal F} = \oplus_{n=0}^\infty {\cal H}_n 
\ee
of the $n$-particle Hilbert spaces ${\cal H}_n$. The pertinent 
mathematical details can be found in Refs. [41,42,44,63]. Define a 
restriction of the operator $\hat A$ on ${\cal H}_n$ as $\hat A_n$.
Then the statistical state in the canonical ensemble can be denoted 
as $<{\cal A}_N>_\rho$, with a fixed density $\rho$ and the number 
of particles $N$. In view of the structure (108), the states 
$<{\cal A}>_\mu$ and $<{\cal A}_N>_\rho$ are related through the 
integral
\be
\label{109}
<{\cal A}>_{\mu(\rho)} \; = \int_0^\infty K(\rho,x) 
<{\cal A}_{N(x)}>_\rho \; dx \; ,
\ee
in which $\mu=\mu(\rho)$ is a solution of Eq. (107) and $N(x)\equiv xV$. 
The kernel $K(\rho,x)$ is called the Kac density. The corresponding 
states coincide, when in the thermodynamic limit
$$
K(\rho,x)\; \ra \; \dlt(\rho-x) \; .
$$
Then one has
\be
\label{110}
<{\cal A}>_{\mu(\rho)} \; = \; <{\cal A}_N>_\rho \; ,
\ee
which signifies the {\it statistical equivalence} of grand canonical 
and canonical ensembles. 

Comparing the statistical states, one has to be very cautious, remembering 
that it may happen that there is not just the sole canonical or grand 
canonical ensemble, but there could be several such ensembles depending on 
additional constraints specifying the properties of the considered system. 
This is related to the nonuniqueness of the Fock space (108) and the 
existence of nonequivalent operator representations, as is discussed in 
Sections VII and VIII. Therefore, one has, first of all, to define the 
appropriate representative ensembles and only after this one can compare 
the related averages. If at least one of the ensembles is not representative, 
then there is no sense to compare the averages and equality (110) does not 
need to be valid.

As an example, let us take a Bose-condensed system, which, according 
to the previous sections, can be considered either using an operator 
representation on the gauge-symmetric space ${\cal F}(\psi)$ or 
employing the Bogolubov representation on the space ${\cal F}(\psi_1)$, 
with broken gauge symmetry. In the former case, some fictitious 
instabilities may arise and Eq. (110) may become invalid. However, 
this would not imply nonequivalence of the ensembles, but would simply 
mean that nonrepresentative ensembles are involved.

Recall as well that a representative ensemble is assumed to represent 
a stable system. For unstable models, Eq. (110) does not have to be 
always valid. For instance, if we consider the ideal Bose gas in a 
box, which, as has been explained above, is not stable, then there is 
no reason to require that Eq. (110) be true. This is really so below 
the condensation point [64,65], where the Bose-condensed gas becomes 
unstable. This instability is manifested by thermodynamically anomalous 
density fluctuations. The ideal Bose gas is also shown [65] to be 
unstable with respect to boundary conditions, whose slight variation 
leads to a dramatic change of the spatial particle distribution, 
even in the thermodynamic limit. This is contrary to the behaviour 
of realistic stable systems, for which the influence of boundary 
conditions disappears in the thermodynamic limit. Changing, for the 
ideal Bose gas, the boundary conditions from repulsive to attractive 
[65] transforms the Bose-Einstein condensation from the bulk phenomenon 
to a strange surface effect, when the condensate is localized in a 
narrow domain in the vicinity of the system surface, being mainly 
concentrated at the corners of an infinite box. It is clear that a 
system, in which the condensate is localized somewhere at the corners 
of an infinite volume, is a rather unphysical object.

Thus, formally comparing two ensembles, one sometimes can arrive at their 
seeming nonequivalence. This, however, in no way invalidates the basic 
principle of statistical mechanics stating the ensemble equivalence. This 
just means that at least one of the compared ensembles is not representative, 
which also includes that the system may be intrinsically unstable. {\it The 
principle of equivalence holds only for representative ensembles, which 
represent stable systems}.

\section{Conclusion}

The analysis is given of the relation between the stability properties
of statistical systems and the fluctuations of observable quantities. The
emphasis is made on the composite observables that are represented by the
sums of several terms. The main result of the paper is the theorem connecting
the global fluctuations of an observable with the partial fluctuations of its
components. The theorem is general, being formulated for an arbitrary operator
represented as a sum of linearly independent self-adjoint operators. These
operators can be associated with the total and partial observable quantities
of a statistical system. The theorem tells that: The total dispersion of an
operator, being a sum of linearly independent self-adjoint operators, is 
thermodynamically anomalous if and only if at least one of the partial 
dispersions is anomalous, with the power of $N$ in the total dispersion 
defined by the largest partial dispersion. Conversely, the total dispersion 
is thermodynamically normal if and only if all partial dispersions are normal.

The theorem allows us to understand the relation between the fluctuations of 
partial observables and the fluctuations of the total observable. Respectively,
the character of partial fluctuations turns out to be directly related to the 
stability of statistical systems. Several examples illustrate the practicality
of the theorem, helping to avoid wrong conclusions that could happen when 
studying the behaviour of partial observables. In particular, the fluctuations
of condensed, as well as noncondensed particles, in a Bose-condensed system 
must be normal, if the system is assumed to be stable. In the same way, 
fluctuations is systems with continuous symmetry are also thermodynamically 
normal.

Breaking of gauge symmetry helps to eliminate fictitious instabilities arising
in Bose-condensed systems. Generally, it is crucially important that a system 
be characterized by its representative ensemble. This makes it possible to 
avoid artificial contradictions in the theory and the related unphysical 
instabilities. One of the basic principles of statistical mechanics, the 
principle of ensembles equivalence, holds only for representative ensembles 
correctly representing stable statistical systems.

\vskip 5mm

{\bf Acknowledgement}

\vskip 2mm

I appreciate the Mercator Professorship of the German Research Foundation.
I am indebted to P. H\"anggi for highly useful advice and several constructive 
remarks.

\newpage

{\large{\bf Appendix A. Noncommutativity of Bogolubov Shift}}

\vskip 5mm

This Appendix illustrates the noncommutativity of the Bogolubov shift
and the Hartree-Fock-Bogolubov approximation (HFB approximation). When one
accomplishes in function (52), first, the Bogolubov shift (53) and then the
HFB approximation for $\psi_1(\br)$, one gets expression (55), with the
correct limiting behaviour. But in the other way round, employing, first, the
HFB approximation for $\psi(\br)$ and, after this, substituting the Bogolubov
shift (53), one gets
$$
g(\br_{12}) = 1 + \frac{2\rho_0^2}{\rho} + \frac{2\rho_0}{\rho^2} \;
{\rm Re}\left [ \rho_1(\br_1,\br_2) + \sgm_1(\br_1,\br_2)\right ] +
\frac{1}{\rho^2}\left \{ |\rho_1(\br_1,\br_2)|^2 +  |\sgm_1(\br_1,\br_2)|^2
\right \} \; .
$$
The limiting behaviour of this pair correlation function is not correct,
since here
$$
\lim_{\br_{12}\ra\infty} g(\br_{12}) =  1 + \frac{2\rho_0^2}{\rho^2} \; ,
$$
which would be true only when $\rho_0\equiv 0$. But when $\rho_0\neq 0$, we
confront the problem of the condensate overcounting. Thence, these procedures
are not commutable. And one has, first, to introduce the Bogolubov shift
(53) and only after this to resort to the HFB approximation.

\newpage

{\large{\bf Appendix B. Nonexistence of Phase Operator}}

To show that the representation (63) does not exist, we may use the method 
of reduction to absurdity. Suppose that this representation is correct. 
Then, from the commutation relation
$$
\left [ \hat n(\br),\; \hat\vp(\br')\right ] = i\dlt(\br-\br')\; ,
$$
we obtain for the number-of-particle operator
$$
\hat N \equiv \int \hat n(\br)\; d\br
$$
the commutaton relation
$$
\left [ \hat N, \; \hat\vp(\br) \right ] = i \; .
$$
From here, taking the matrix element with respect to the number basis 
$\{|n>\}$, for which $\hat N|n>=n|n>$, we find
$$
(n-n') < n|\vp(\br)|n'> \; = i\dlt_{nn'} \; .
$$
Setting here $n=n'$, we get the senseless equality $i=0$. Thus, the 
representation (63) does not exist.

Now, suppose that the representation (64) is correct. Then for the 
density operator, we have
$$
\hat n(\br) \equiv \psi^\dgr(\br) \psi(\br) = \rho(\br) \; .
$$
Hence, the number-of-particle operator becomes identical to the total 
number of particles,
$$
\hat N = \int \rho(\br)\; d\br = N \; .
$$
At the same time, there is an exact relation
$$
\left [ \psi(\br),\hat N\right ] =\psi(\br) \; .
$$
Using this for $\hat N=N$, we get the senseless equality $\psi(\br)=0$. 
Hence, the representation (64) is wrong.

In this way, neither representation (63) nor representation (64) are 
correct. The phase operator, defined through these representations, does 
not exist. To introduce correctly a kind of a quasi-phase operator, one 
should employ the Pegg-Barnett technique [36].

\vskip 5mm

\newpage

{\large{\bf Appendix C. Gauge-Symmetry Breaking}}

\vskip 5mm

The simple method of infinitesimal sources may not always break 
gauge symmetry. To illustrate this, it is sufficient to give at least 
one counterexample. For this purpose, let us consider the Hamiltonian
$$
H = \int \psi^\dgr(\br)\om(\br)\psi(\br)\; d\br \; ,
$$
with a positive function $\om(\br)>0$. This Hamiltonian is invariant under 
the gauge transformation
$$
\psi (\br) \; \longrightarrow \; e^{i\al} \psi(\br) \; ,
$$
where $\al$ is any real-valued number. Hence $<\psi(\br)>=0$. To break the 
gauge symmetry, following the standard method of infinitesimal sources, one 
adds to the Hamiltonian $H$ a term lifting the symmetry. For instance, the 
Hamiltonian
$$
H_\ep \equiv H - \ep \int \left [ \lbd^*(\br) \psi(\br) + \lbd(\br)
\psi^\dgr(\br) \right ] \; d\br \; ,
$$
where $\lbd(\br)$ is a complex-valued function, is not gauge invariant. 
The latter Hamiltonian can be diagonalized by means of the canonical 
transformation
$$
\psi(\br) = \ep \; \frac{\lbd(\br)}{\om(\br)} + \overline\psi(\br) \; ,
$$
in which the new field operator $\overline\psi(\br)$ enjoys the same 
commutation relations as $\psi(\br)$. Then we have
$$
H_\ep =  E_\ep + \int {\overline\psi}^\dgr(\br)
\om(\br)\overline\psi(\br) \; d\br \; ,
$$
with the notation
$$
E_\ep \equiv -\ep^2 \; \int \; 
\frac{|\lbd(\br)|^2}{\om(\br)} \; d\br \; .
$$
For the diagonal in $\overline\psi(\br)$ Hamiltonian $H_\ep$, one has 
$<\overline\psi(\br)>=0$. Therefore
$$
< \psi(\br)> \; = \; \ep \; \frac{\lbd(\br)}{\om(\br)} \; .
$$
According to the method of infinitesimal sources, after calculating the 
averages, one should set $\ep\ra 0$. But then
$$
<\psi(\br)> \; \ra \; 0 \qquad (\ep \ra 0) \; ,
$$
because of which the gauge symmetry has not been broken. Contrary to 
this, the Bogolubov shift (67) is a sufficient condition for gauge-symmetry 
breaking.

\end{document}